\documentclass[10pt, a4paper]{article}

\usepackage[T1]{fontenc}
\usepackage[utf8]{inputenc}
\usepackage{microtype}
\usepackage{lmodern}
\usepackage[english]{babel}

\usepackage{amscd,amsmath,amsfonts,amssymb,amsthm}
\usepackage{bm}
\usepackage{braket}
\usepackage[mathcal]{euscript}
\usepackage{mathrsfs}
\usepackage{mathtools}
\usepackage{tensor}
\usepackage[output-decimal-marker={.}]{siunitx}
\usepackage[only,llbracket,rrbracket]{stmaryrd}
\usepackage{upgreek}

\usepackage{booktabs}
\usepackage{caption}
\usepackage{bmpsize}
\usepackage{multirow}
\usepackage{subfig}
\usepackage{tabularx}
\usepackage[svgnames,dvipsnames,table]{xcolor}
\usepackage{enumerate}

\date{}

\begin{document}

\title{\bf Platonic Localisation: One Ring to Bind Them}

\author{A.B. Movchan$^{a}$, R.C. McPhedran$^{b}$, G. Carta$^{c,*}$, R.V. Craster$^{d}$ \\
\small{$^a$ Department of Mathematical Sciences, University of Liverpool, Liverpool, UK} \\
\small{$^b$ School of Physics, University of Sydney, Sydney, Australia} \\
\small{$^c$ Mechanical Engineering and Materials Research Centre,} \\
\small{Liverpool John Moores University, Liverpool, UK} \\
\small{$^d$ Department of Mathematics, Imperial College, London, UK} \\
\small{$^*$Corresponding author; email address: G.Carta@ljmu.ac.uk} }

\maketitle

\begin{abstract}
\noindent
In this paper, we present an asymptotic model describing localised flexural vibrations along a structured ring containing point masses or spring-mass resonators in an elastic plate. The values for the required masses and stiffnesses of resonators are derived in a closed analytical form. It is shown that spring-mass resonators can be tuned to produce a ``negative inertia'' input, which is used to enhance localisation of waveforms around the structured ring. Theoretical findings are accompanied by numerical simulations, which show exponentially localised and leaky modes for different frequency regimes.
\end{abstract}


\section{Introduction}
\label{Introduction}

The interest in wave propagation and localised waveforms in structured media has led to analysis of defects in crystalline solids, dynamic Green's kernels in lattices, transmission resonances for grating stacks, and novel designs of structured  waveguides.  While motivated by practical applications in acoustics and optics, these problems bring new challenges in elasticity: in particular, structured flexural elastic plates incorporate effects attributed to the interaction between evanescent and propagating modes. In turn, this may lead to localised flexural vibration modes within finite structured clusters. Lattice dynamics and analysis of point defects were extensively studied in \cite{Mar1, Mar2}. Analysis of propagating modes initiated by point defects was published in \cite{Martin}, and stop band Green's functions representing exponentially localised waveforms were studied in \cite{Movchan_Slepyan}.
A scalar problem governed by the Helmholtz operator for a circular cluster of small rigid disc-shaped inclusions was analysed in \cite{Martin_PRSA}.
Localisation and waveguiding of flexural waves by structured semi-infinite grating stacks were studied in \cite{Hasl_s-inf1, Hasl_s-inf2}, and transmission resonances for flexural waves in an elastic plate containing gratings of rigid pins were analysed in \cite{Hasl_edit}.

\begin{figure}%[!htcb]
\centering
\includegraphics[width=0.5\columnwidth]{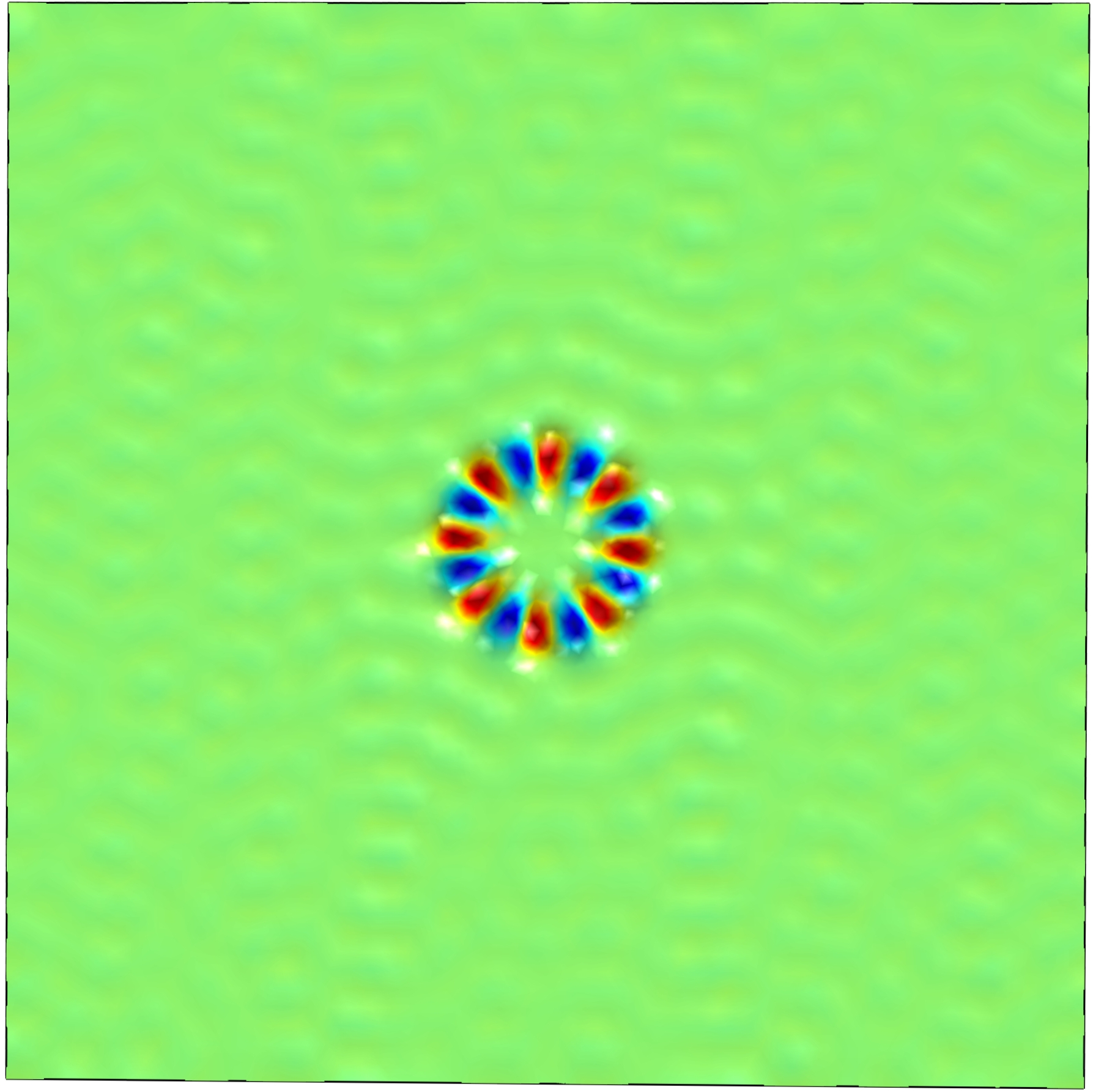}
\caption{\footnotesize Illustration of an exponentially localised mode in an elastic plate, incorporating a circular ring of spring-mass resonators. In the computations, $64$ resonators are situated along a circle of radius $R = 1$ m, and each of them possesses a mass $\mathcal{m} = 0.5$ kg and a stiffness $c = 346.476$ kN/m. The plate is made of aluminum and presents a square shape, with a side length $L = 10$ m and a thickness $h = 0.005$ m.}
\label{FigureExpLocalisedModesIntro}
\end{figure}

In the present paper, we develop an asymptotic homogenisation model for evaluation of physical parameters associated with localised flexural waveforms, which occur around circular clusters of point masses or spring-mass resonators embedded in elastic plates. An example of such a localised waveform is shown in Fig. \ref{FigureExpLocalisedModesIntro}. Often, ``leaky waves'' can be observed near clusters of defects such as point masses, rigid pins or spring-mass resonators. It is also noted that the rate of localisation can be enhanced by using spring-mass resonators with the masses and stiffness chosen according to the procedure discussed in the text below.
In particular, Fig. \ref{FigureExpLocalisedModesIntro} shows the results of computations for a ring of specially designed spring-mass resonators.

%%%%%%%%%%%%%%%%%%%%%%%%%%%%%%%%%%%%%%%%%%%%
%%%%%%%%%%%%%%%%%%%%%%%%%%%%%%%%%%%%%%%%%%%%

The recent papers \cite{MalingCraster2016,Maling2017} present an asymptotic analysis of whispering Bloch modes around a circular array of inclusions for acoustic or electromagnetic waves. The results have been obtained for the case of the Helmholtz operator and the Neumann boundary conditions set at the boundaries of the scatterers. The thesis \cite{Hugo2016} has addressed flexural wave localisation, and waveguiding by clusters of point masses inserted in an elastic plate.
Clusters of resonators including springs and masses, embedded into an elastic plate, were studied in \cite{Hasl_s-inf1} in the context of wave scattering and trapping by semi-infinite structured gratings.

The asymptotic model developed in the present paper reduces the formulation to a spectral problem for an integral operator. The solution is obtained in a closed analytical form. Special attention is given to localised waveforms in a structured ring associated with spring-mass resonators of negative inertia, a condition we show to be necessary for their occurrence in the finite cluster represented by the structured ring.
%Leaky waveforms occur when the inertia is positive.

The structure of the paper is as follows: a Green's function formulation for vibrations of point masses in a flexural plate is introduced in Section \ref{GreenfunctionSection}. Section \ref{HomogenisationSection} includes the homogenisation approximation and reduction to the spectral problem for an integral operator. Normalisation and closed form evaluation of certain integrals for the special case of a circular cluster are discussed in Sections \ref{NormalisationSection} and \ref{ClosedformSection}. Localised waveforms for different types of resonators are described in Sections \ref{FrequencieslocalisedmodesSection} and \ref{ResonatorsSection}. Flexural wave localisation in an infinite grating of point masses is discussed in Section \ref{InfiniteGratingSection}. Finally, Section \ref{ConcludingSection} includes concluding remarks and discussion.

%%%%%%%%%%%%%%%%%%%%%%%%%%%%%%%%%%%%%%%%%%%%
\begin{figure}%[!htcb]
	\centering
	\includegraphics[width=0.5\columnwidth]{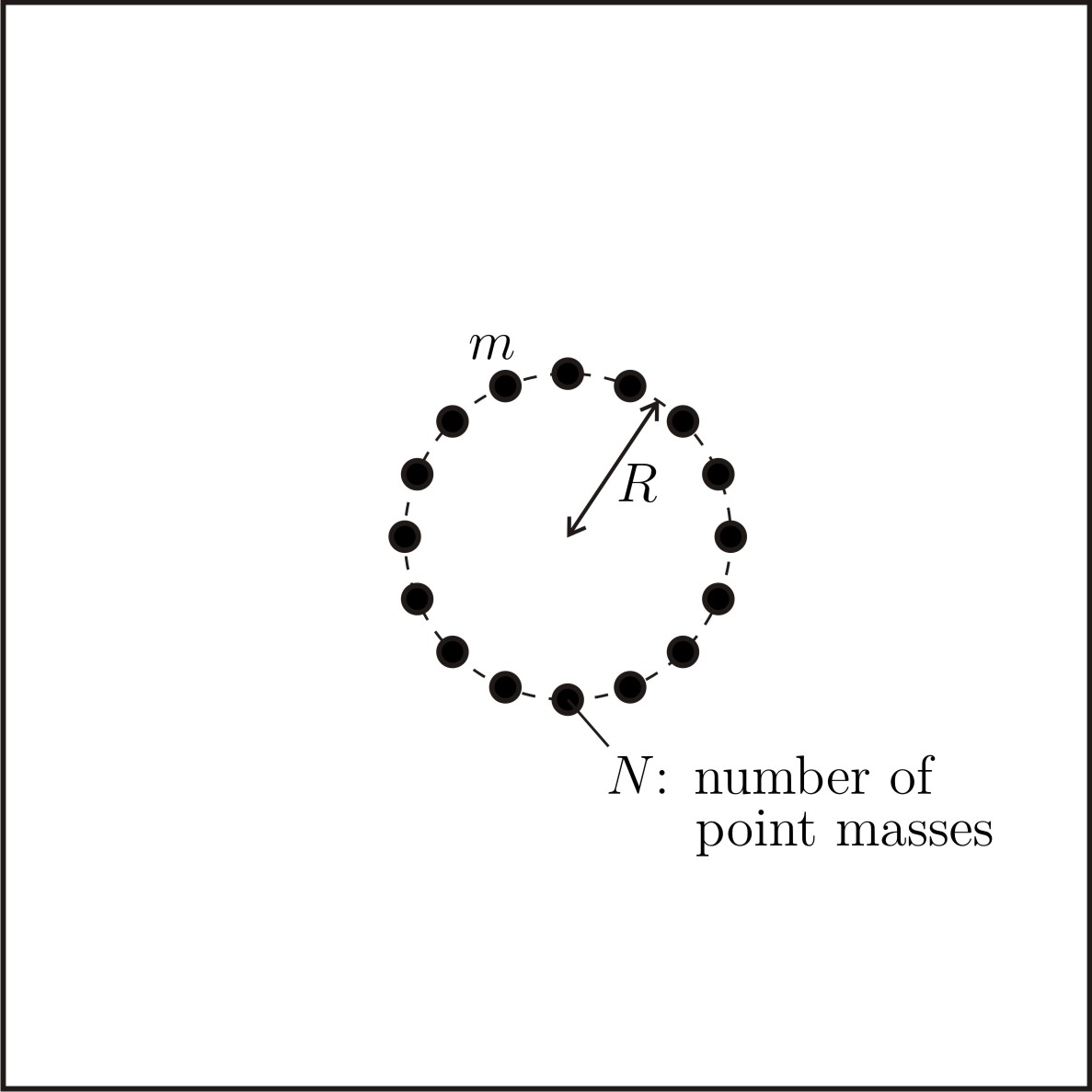}
	\caption{\footnotesize Ring of point masses in an elastic plate of thickness $h$ and flexural stiffness $D$.}
	\label{FigureModel}
\end{figure}
%%%%%%%%%%%%%%%%%%%%%%%%%%%%%%%%%%%%%%%%%%%%

\section{Green's function and vibration of an inertial cluster}
\label{GreenfunctionSection}

The problem is formulated in terms of the dynamic Green function $G(r;\beta)$, $r = \left|\bm{x}\right|$, which satisfies the Kirchhoff-Love equation governing time-harmonic flexural waves (see, for example, \cite{Graff1991}) associated with a point source at the origin:
\begin{equation}\label{EquationGreenFunction}
\Delta^{2} G(r;\beta) - \beta^{4} G(r;\beta) + \delta(\bm{x}) = 0 \, .
\end{equation}
Here, the spectral parameter $\beta$ and the radian frequency $\omega$ are related by
\begin{equation}
\beta^{4} = \frac{\rho h \omega^{2}}{D}, \label{betom}
\end{equation}
with $\rho$, $h$ and $D$ being the mass density, the thickness and the flexural rigidity of the plate, respectively.
The explicit form for $G(r;\beta)$ is given in terms of Bessel functions (see, for example, \cite{eandp2007,us2007})
\begin{equation}\label{GreenFunction}
G(r;\beta) = -\frac{\mathrm{i}}{8 \beta^{2}} \left\{ \mathrm{H}_{0}^{(1)}(\beta r) + \frac{2 \mathrm{i}}{\pi} \mathrm{K}_{0}(\beta r) \right\} \, .
\end{equation}
Note that $G(r; \beta)$ is bounded as $r\rightarrow 0$, which is important for our model.

For a ring of $N$ points, each with mass $m$, shown in Fig. \ref{FigureModel},  we consider a time-harmonic flexural displacement $u(\bm{x}) e^{-i \omega t}$, where $u(\bm{x})$ satisfies the following equation governed by the bi-harmonic operator
\begin{equation}\label{EquationPlate}
\Delta^{2} u - \frac{\rho h \omega^{2}}{D} u - \frac{m \omega^{2}}{D} \sum_{j=1}^{N} u\left(\bm{a}^{(j)}\right) \delta\left(\bm{x}-\bm{a}^{(j)}\right) = 0 \, ,
\end{equation}
where $\bm{a}^{(j)}, ~ j = 1,\ldots,N,$ represent the positions of the point masses within the structured ring. Here $\left|\bm{a}^{(j)}\right|=R$, where $R$ is the radius of the ring.
The solution $u(\bm{x})$ satisfies the following relation
\begin{equation}\label{DisplacementPlate}
u(\bm{x}) = - \frac{m \omega^{2}}{D} \sum_{j=1}^{N} u\left(\bm{a}^{(j)}\right) G\left(\left|\bm{x}-\bm{a}^{(j)}\right|; \beta \right) \, .
\end{equation}
In particular, when $\bm{x} = \bm{a}^{(k)}$ we deduce
\begin{equation}\label{DisplacementSystem}
u(\bm{a}^{(k)}) = - \frac{m \omega^{2}}{D} \sum_{j=1}^{N} u\left(\bm{a}^{(j)}\right) G\left(\left|\bm{a}^{(k)}-\bm{a}^{(j)}\right|; \beta \right) \, , \qquad k = 1,\ldots,N.
\end{equation}
By looking for a non-trivial solution, incorporating complex $\beta$ and $u(\bm{a}^{(k)}), k = 1,\ldots,N,$ (see, for example \cite{Hugo2016}), one will obtain a description of a ``leaky wave'' around a cluster of point masses, and the imaginary part of $\beta$ characterises the ``leakage'' to infinity.
When $N$ becomes large, we propose a homogenisation model and  find its solution in a simple analytical form.
In particular, it is shown that for the case of a ``negative inertia'' of the structured ring, real roots $\beta$ are identified for the homogenisation model. It is also shown that the effect of negative inertia is achieved by using spring-mass resonators instead of point masses.

\section{Homogenisation approximation}
\label{HomogenisationSection}

For the circle of radius $R$ centered at the origin we use the notations
\begin{equation}\label{FormulaeCircle}
\bm{x} = R \mathrm{e}^{\mathrm{i}\theta} \, , \; \bm{a}^{(j)} = R \mathrm{e}^{\mathrm{i}\theta_{j}} \, , \; \theta_{j} = \frac{2 \pi}{N}(j-1) \, , \qquad 1 \le j \le N \, .
\end{equation}
For a periodic solution $u(\bm{x})$ of (\ref{DisplacementSystem}), the displacements of the point masses are $u\left(\bm{a}^{(j)}\right) = U \mathrm{e}^{\mathrm{i} k \theta_{j}}$, where $k$ is an integer. The system of equations  \eqref{DisplacementSystem} can be rewritten in the form:
\begin{equation}\label{DisplacementPlate2}
\frac{N}{2 \pi R} \sum_{j=1}^{N} u\left(\bm{a}^{(j)}\right) G\left(\left|\bm{a}^{(p)}-\bm{a}^{(j)}\right|; \beta \right) \frac{2 \pi R}{N} + \frac{D}{m \omega^{2}} u\left(\bm{a}^{(p)}\right) = 0 \, ,
\end{equation}
where $M = N m$ is the ``total mass'' of the ring.

\subsection{Spectral problem for an integral operator}
\label{intop}

Assuming that $M$ is constant and taking the limit as $N\rightarrow \infty$, we deduce the homogenised spectral problem
\begin{equation}\label{Plate3}
\mathrm{L} u + \Lambda u = 0 \, ,
\end{equation}
where
\begin{equation}\label{DisplacementPlate3L}
\mathrm{L} u = \int\displaylimits_{\left\{ \left| \bm{Y} \right| = R \right\}} u(\bm{Y}) G(\left|\bm{Y}-\bm{X}\right|;\beta) \mathrm{d}l
\end{equation}
and
\begin{equation}\label{DisplacementPlate3Lambda}
\Lambda = \lim_{N \to \infty} \frac{N D}{M \omega^{2}} \frac{2 \pi R}{N} = \frac{2 \pi R D}{M \omega^{2}} \, .
\end{equation}

To solve \eqref{Plate3}, we seek $u$ in the form $u(R \mathrm{e}^{\mathrm{i}\theta}) = U \mathrm{e}^{\mathrm{i} k \theta}$, where $k$ is an integer and $0 \le \theta < 2 \pi$. Then the equation takes the form
\begin{equation}\label{DisplacementPlate4}
\int_{0}^{2 \pi} \mathrm{e}^{\mathrm{i} k \theta_{Y}} G\left( \left| R (\mathrm{e}^{\mathrm{i} \theta_{Y}}-\mathrm{e}^{\mathrm{i} \theta_{X}}) \right|; \beta \right) \mathrm{e}^{-\mathrm{i} k \theta_{X}} \mathrm{d}\theta_{Y} + \frac{2 \pi D}{M \omega^{2}} = 0 \, .
\end{equation}
Note that $\left| R (\mathrm{e}^{\mathrm{i} \theta_{Y}}-\mathrm{e}^{\mathrm{i} \theta_{X}}) \right| = \left| R (\mathrm{e}^{\mathrm{i} (\theta_{Y}-\theta_{X})} - 1) \right|$, and hence
\begin{equation}\label{DisplacementPlate5}
\int_{0}^{2 \pi} \mathrm{e}^{\mathrm{i} k \theta} G\left( \left| R (\mathrm{e}^{\mathrm{i} \theta}-1) \right|; \beta \right) \mathrm{d}\theta + \frac{2 \pi D}{M \omega^{2}} = 0 \, ,
\end{equation}
where $\mathrm{d}\theta = \mathrm{d}l / R$. We also note that $\left| R (\mathrm{e}^{\mathrm{i} \theta}-1) \right| = 2 R \left| \sin{(\theta/2)} \right|$. Hence
\begin{equation}\label{DisplacementPlate6}
\int_{0}^{2 \pi} \mathrm{e}^{\mathrm{i} k \theta} G\left( 2 R \left| \sin{\left(\frac{\theta}{2}\right)} \right|; \beta \right) \mathrm{d}\theta + \frac{2 \pi \rho h}{M \beta^{4}} = 0 \, ,
\end{equation}
from which we can determine the ``eigenvalue'' $\beta$.

\subsection{Normalisation}
\label{NormalisationSection}

Eq. \eqref{DisplacementPlate6} is normalised  by introducing the non-dimensional spectral parameter $\tilde{\beta} = 2 \beta R$. Taking into account that the Green's function satisfies the relation
\begin{equation}\label{GreenFunctionNormalised}
G \left( 2 R \sin{\left(\frac{\theta}{2}\right)};\beta \right) = 4 R^{2} G \left( \sin{\left(\frac{\theta}{2}\right)};\tilde{\beta} \right) \, ,
\end{equation}
the normalised equation is written in the form
\begin{equation}\label{DisplacementPlateNormalised}
\tilde{\beta}^{4} \int_{0}^{\pi} \cos{(k \theta)} G\left(\sin{\left(\frac{\theta}{2}\right)}; \tilde{\beta} \right) \mathrm{d}\theta + 4 \pi \rho h \frac{R^{2}}{M} = 0 \, .
\end{equation}
This equation is solved exactly and the evaluation of the integral term in (\ref{DisplacementPlateNormalised}) is discussed in the next section.

\subsection{Evaluation of the integral in (\ref{DisplacementPlateNormalised}) }
\label{ClosedformSection}

The integral on the left-hand side of Eq. \eqref{DisplacementPlateNormalised} has the following closed form expression:
\begin{equation}\label{ClosedFormSolutionIntegral}
\begin{split}
\tilde{\beta}^{4} \int_{0}^{\pi} & \cos{(n \theta)} G\left(\sin{\left(\frac{\theta}{2}\right)}; \tilde{\beta} \right) \mathrm{d}\theta = \frac{\pi \tilde{\beta}^{2}}{8} \left\{ -\mathrm{i} \, \mathrm{J}_{n}^{2}\left(\frac{\tilde{\beta}}{2}\right) \right. \\
&\left. + \mathrm{J}_{n}\left(\frac{\tilde{\beta}}{2}\right) \mathrm{Y}_{n}\left(\frac{\tilde{\beta}}{2}\right) + \frac{2}{\pi} \mathrm{I}_{n}\left(\frac{\tilde{\beta}}{2}\right) \mathrm{K}_{n}\left(\frac{\tilde{\beta}}{2}\right) \right\} \, ,
\end{split}
\end{equation}
where $n$ is an integer, $\mathrm{J}_{n}$ and $\mathrm{Y}_{n}$ are Bessel's functions, while $\mathrm{I}_{n}$ and $\mathrm{K}_{n}$ are modified Bessel's functions.

Derivation of the above formula is based on the evaluation of integrals related to addition theorems for cylinder functions discussed in Chapter XI of Watson \cite{Watson1962}, and also linked to the classical work by Gegenbauer \cite{Gegenbauer1875}.

Namely, we have
\begin{eqnarray}
&&\int_0^{\pi/2} \cos (2  n\phi ) J_0(\tilde{\beta} \sin \phi) d\phi=\int_0^{\pi/2} \cos (2  n\phi ) J_0(\tilde{\beta} \cos \phi) d\phi \nonumber \\
&=&\frac{ \pi }{2} J_n^2\left( \frac{{\tilde \beta}}{2}\right)
\label{t1int}
\end{eqnarray}
and
\begin{eqnarray}
&&\int_0^{\pi/2} \cos (2  n\phi ) Y_0(\tilde{\beta} \sin \phi) d\phi=\int_0^{\pi/2} \cos (2  n\phi ) Y_0(\tilde{\beta} \cos \phi) d\phi \nonumber \\
&=&\frac{ \pi }{2} J_n\left( \frac{{\tilde \beta}}{2}\right) Y_n\left( \frac{{\tilde \beta}}{2}\right).
\label{SGint2}
\end{eqnarray}
We also get
\begin{eqnarray}
&&\int_0^{\pi/2} \cos (2  n\phi ) H_0^{(1)}(\tilde{\beta} \sin \phi) d\phi=\int_0^{\pi/2} \cos (2  n\phi ) H_0^{(1)}(\tilde{\beta} \cos \phi) d\phi \nonumber \\
&=&\frac{ \pi }{2} J_n\left( \frac{{\tilde \beta}}{2}\right) H_n^{(1)}\left( \frac{{\tilde \beta}}{2}\right).
\label{SGint3}
\end{eqnarray}
Replacing real Bessel function arguments by imaginary arguments, we further obtain
\begin{eqnarray}
&&\int_0^{\pi/2} \cos (2  n\phi ) K_0(\tilde{\beta} \sin \phi) d\phi=\int_0^{\pi/2} \cos (2  n\phi ) K_0(\tilde{\beta} \cos \phi) d\phi \nonumber \\
&=&\frac{ \pi }{2} I_n\left( \frac{{\tilde \beta}}{2}\right) K_n\left( \frac{{\tilde \beta}}{2}\right).
\label{SGint4}
\end{eqnarray}

Combining these integrals (not all of which are given in the standard tables), we obtain the desired closed-form expression (\ref{ClosedFormSolutionIntegral}).

%%%%%%%%%%%%%%%%%%%%%%%%%%%%%%%%%%%%%%%%%%%%
\begin{figure}%[!htcb]
\centering
\includegraphics[width=1.0\columnwidth]{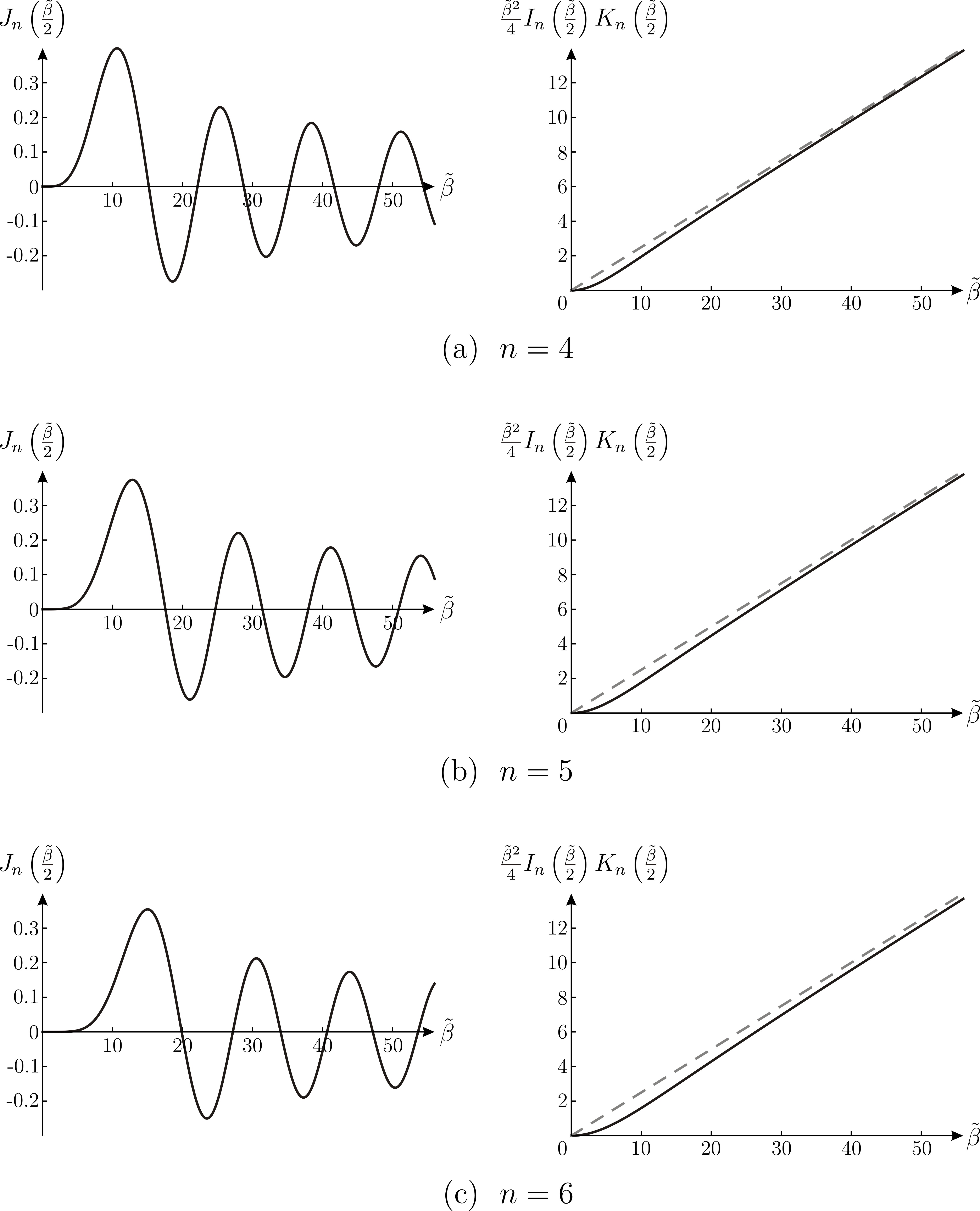}
\caption{\footnotesize Graphs of the functions $J_n(\tilde{\beta}/2)$ and $\tilde{\beta}^2/4 \; I_n(\tilde{\beta}/2) K_n(\tilde{\beta}/2)$ versus $\tilde{\beta}$ for different values of the mode number $n$: (a) $n = 4$, (b) $n = 5$, (c) $n = 6$.}
\label{FigureFunctions}
\end{figure}
%%%%%%%%%%%%%%%%%%%%%%%%%%%%%%%%%%%%%%%%%%%%

\section{Frequencies of the localised modes}
\label{FrequencieslocalisedmodesSection}

Localised modes in the ring of masses are obtained as follows:
\begin{itemize}
\item First, the normalised frequency parameter is determined by solving the equation corresponding to the imaginary term in (\ref{ClosedFormSolutionIntegral}):
\begin{equation}\label{ZeroImaginaryPart}
\mathrm{J}_{n}\left(\frac{\tilde{\beta}}{2}\right) = 0 \, .
\end{equation}
\item Then, given $\tilde{\beta}$, the ring radius $R$ and the total mass $M$ are derived from the equation
\begin{equation}\label{ZeroRealPart}
\frac{\tilde{\beta}^{2}}{4}\mathrm{I}_{n}\left(\frac{\tilde{\beta}}{2}\right) \mathrm{K}_{n}\left(\frac{\tilde{\beta}}{2}\right) + 4 \pi \rho h \frac{R^{2}}{M} = 0 \, .
\end{equation}
\end{itemize}
Note that there are real solutions $\tilde{\beta}$ only if $M < 0$. For positive total mass, the system exhibits only leaky modes.

\subsection{Asymptotic simplification of (\ref{ZeroRealPart}) for large $\tilde\beta$}
\label{AsymptoticsSection}

Using the following asymptotic approximations for $\tilde{\beta} \gg 1$:
\begin{subequations}\label{AsymptoticApproximations}
\begin{equation}
\mathrm{I}_{n}(z) = \sqrt{\frac{1}{2 \pi z}} \mathrm{e}^{z} \left( 1 + \mathcal{O} \left(\frac{1}{z}\right) \right) \, ,
\end{equation}
\begin{equation}
\mathrm{K}_{n}(z) = \sqrt{\frac{\pi}{2 z}} \mathrm{e}^{-z} \left( 1 + \mathcal{O} \left(\frac{1}{z}\right) \right) \, ,
\end{equation}
\begin{equation}
\mathrm{I}_{n}(z) \mathrm{K}_{n}(z) = \frac{1}{2 z} + \mathcal{O} \left(\frac{1}{z^{2}}\right) \, ,
\end{equation}
\end{subequations}
we derive
\begin{equation}\label{AsymptoticZeroRealPart}
\frac{\tilde{\beta}^2}{4} \mathrm{I}_{n}\left(\frac{\tilde{\beta}}{2}\right) \mathrm{K}_{n}\left(\frac{\tilde{\beta}}{2}\right) \approx \frac{\tilde{\beta}}{4} \, .
\end{equation}
Hence:
\begin{equation}\label{BetaTildeLocalisedMode}
\tilde{\beta} + 16 \pi \rho h \frac{R^{2}}{M} = 0 \, .
\end{equation}

Graphs of $J_n(\tilde{\beta}/2)$ and $\tilde{\beta}^2/4 \; I_n(\tilde{\beta}/2) K_n(\tilde{\beta}/2)$ are given in Fig. \ref{FigureFunctions} for three values of $n$.
Furthermore, the linear asymptotic approximation (\ref{AsymptoticZeroRealPart}) is illustrated in the figure by dashed lines.

%%%%%%%%%%%%%%%%%%%%%%%%%%%%%%%%%%%%%%%%%%%%
\begin{figure}%[!htcb]
\centering
\includegraphics[width=1.0\columnwidth]{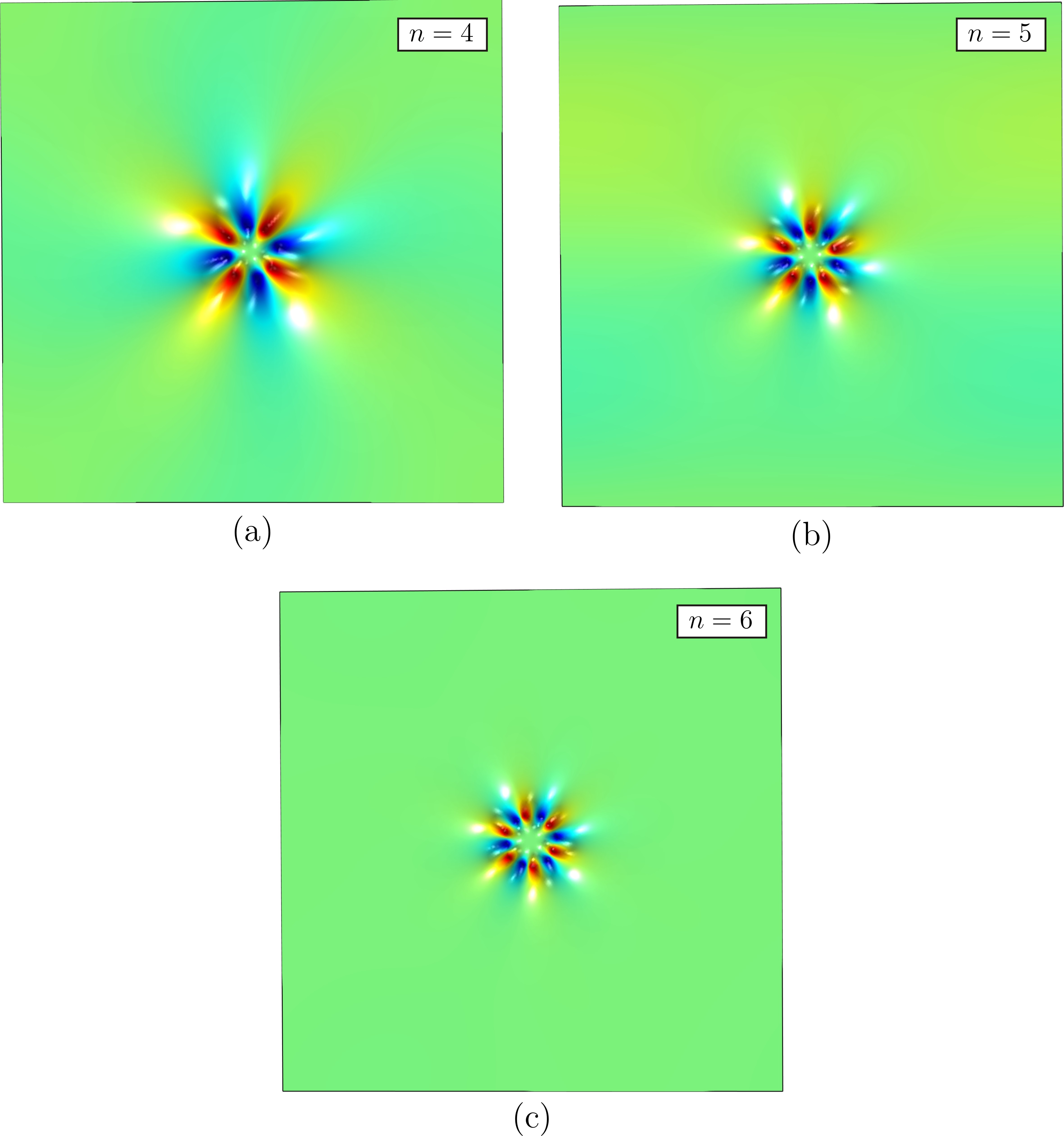}
\caption{\footnotesize Leaky modes in an elastic plate with a ring of masses for different mode numbers $n$, obtained at the following frequencies: (a) $f = 41.89$ Hz, (b) $f = 57.37$ Hz, (c) $f = 73.08$ Hz. In these calculations, each mass is equal to $m = 100$ kg. The values of the frequencies determined analytically are (a) $f = 43.87$ Hz, (b) $f = 62.46$ Hz, (c) $f = 82.30$ Hz. In these simulations, the radius is $R = 0.05$ m and the side length of the square plate is $L = 1$ m, while the other quantities are the same as in Fig. \ref{FigureExpLocalisedModes}. The computations have been performed in \emph{Comsol Multiphysics} (version 5.2a).}
\label{FigureLeakyModes}
\end{figure}
%%%%%%%%%%%%%%%%%%%%%%%%%%%%%%%%%%%%%%%%%%%%

\subsection{Numerical illustrations}
\label{NumIllMasSection}

For positive values of $m$ the equation (\ref{ZeroRealPart}) does not have real solutions, but complex roots can be readily identified.
Hence leaky flexural waveforms may be observed for the case of a ring of point masses $m$ embedded into the elastic plate. Such waves were analysed in detail in \cite{Hugo2016}.

We have constructed illustrative examples of leaky flexural waveforms, as shown in Fig. \ref{FigureLeakyModes},  which are close to the eigenmodes of a finite elastic plate containing a ring of point masses. The values of the physical and geometrical quantities used in the simulations are detailed in the caption of the figure.

The leaky flexural waveforms will be further compared in Section \ref{ResonatorsSection} with exponentially localised modes for spring-mass resonators of negative inertia (compare Figs. \ref{FigureLeakyModes} and \ref{FigureExpLocalisedModes}).

%In this section, we discuss several numerical examples showing localised waveforms along a ring of masses connected to an elastic plate. The numerical simulations presented in this paper have been performed with \emph{Comsol Multiphysics} (version 5.2a).

%%%%%%%%%%%%%%%%%%%%%%%%%%%%%%%%%%%%%%%%%%%%
\begin{figure}[h]%[!htcb]
\centering
\includegraphics[width=0.5\columnwidth]{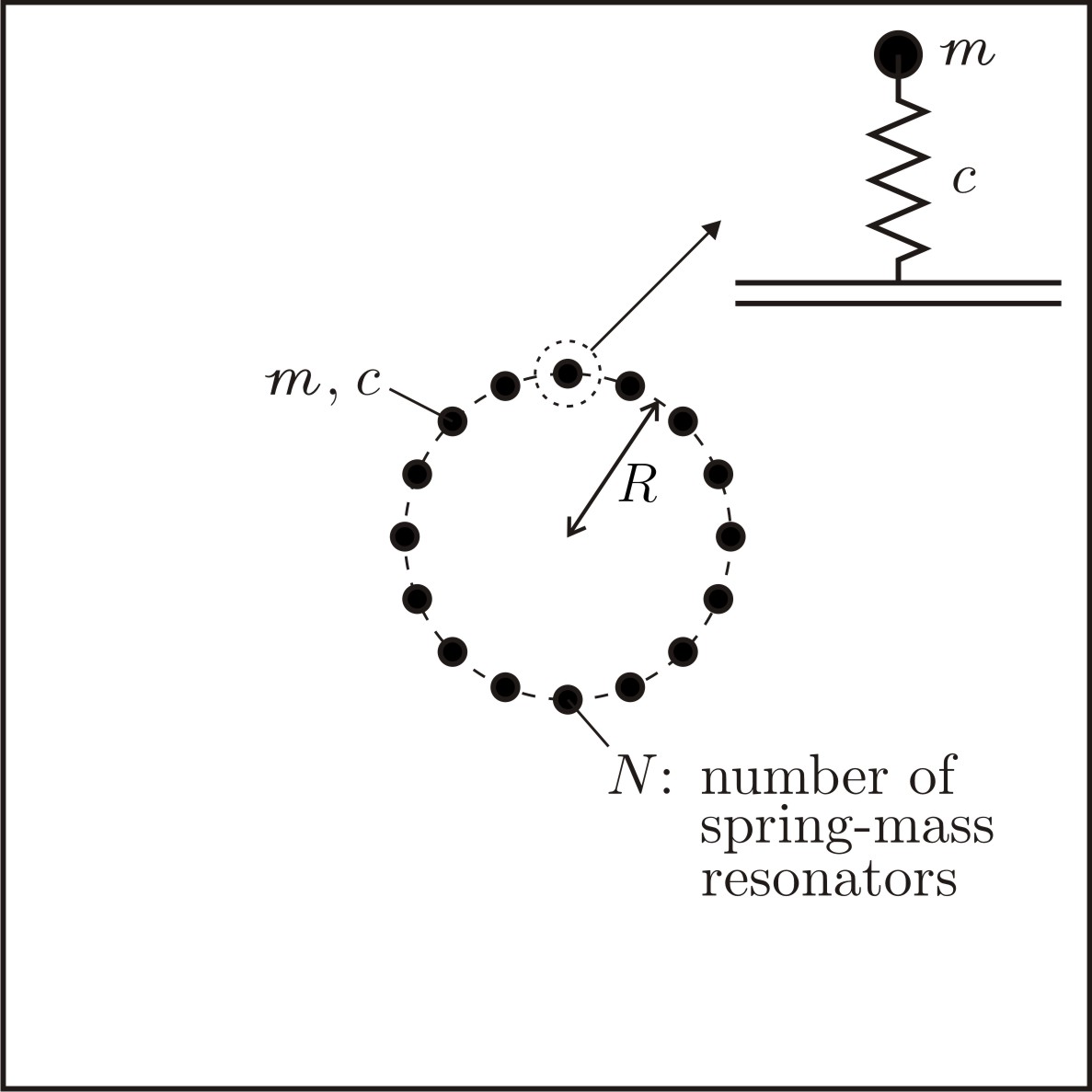}
\caption{\footnotesize Circular cluster of spring-mass resonators attached to an elastic plate.}
\label{FigureModelResonators}
\end{figure}
%%%%%%%%%%%%%%%%%%%%%%%%%%%%%%%%%%%%%%%%%%%%

\section{Resonators with negative inertia}
\label{ResonatorsSection}

Let us consider a plate where the masses are replaced by spring-mass resonators of stiffness $c$ and mass $\mathcal{m}$, as shown in Fig. \ref{FigureModelResonators}. The dynamic response of gratings consisting of spring-mass resonators has been analysed in \cite{Hasl_s-inf1}. Compared to the case of point masses, for the case of spring-mass resonators it has been shown that in the equations of motion the inertia term $m \omega^{2}$ should be replaced by $c \mathcal{m} \omega^{2} / (c - \mathcal{m} \omega^{2})$. Consequently, Eq. \eqref{DisplacementPlate2} becomes
\begin{equation}\label{DisplacementPlate2Resonators}
\frac{N}{2 \pi R} \sum_{j=1}^{N} u\left(\bm{a}^{(j)}\right) G\left(\left|\bm{a}^{(p)}-\bm{a}^{(j)}\right|; \beta \right) \frac{2 \pi R}{N} + \frac{D (c - \mathcal{m} \omega^{2})}{c \mathcal{m} \omega^{2}} u\left(\bm{a}^{(p)}\right) = 0 \, .
\end{equation}
Let $\mathcal{M} = N \mathcal{m}$ be the total mass of resonators and $\mathcal{C} = N c$ be the total stiffness of resonators. Then Eq. \eqref{DisplacementPlate2Resonators} is given by
\begin{equation}\label{DisplacementPlate3Resonators}
\sum_{j=1}^{N} u\left(\bm{a}^{(j)}\right) G\left(\left|\bm{a}^{(p)}-\bm{a}^{(j)}\right|; \beta \right) \frac{2 \pi R}{N} + 2 \pi R \frac{D (\mathcal{C} - \mathcal{M} \omega^{2})}{\mathcal{C} \mathcal{M} \omega^{2}} u\left(\bm{a}^{(p)}\right) = 0 \, .
\end{equation}
The fraction in the last term of \eqref{DisplacementPlate3Resonators} can be chosen to be negative.
In the limit as $N \to \infty$, we have that $\Lambda$ in Eq. \eqref{DisplacementPlate3Resonators} is given by
\begin{equation}\label{DisplacementPlate3LambdaResonators}
\Lambda = 2 \pi R D \frac{\mathcal{C} - \mathcal{M} \omega^{2}}{\mathcal{C} \mathcal{M} \omega^{2}} \, .
\end{equation}

%%%%%%%%%%%%%%%%%%%%%%%%%%%%%%%%%%%%%%%%%%%%
\begin{figure}[!h]%[!htcb]
\centering
\includegraphics[width=1.0\columnwidth]{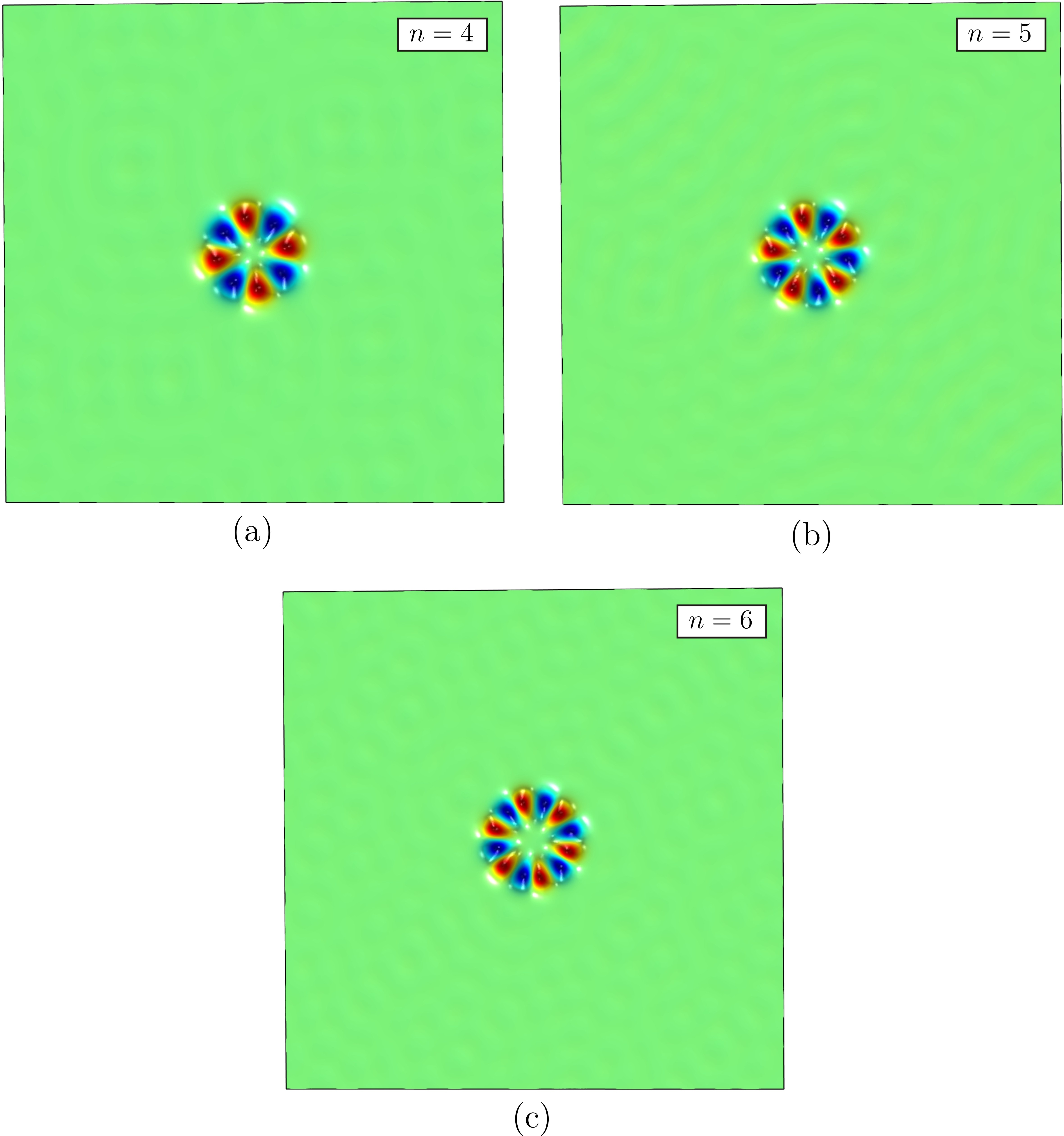}
\caption{\footnotesize Exponentially localised modes along a ring of masses for different mode numbers $n$ and at different frequencies: (a) $f = 70.78$ Hz, (b) $f = 94.32$ Hz, (c) $f = 120.88$ Hz. The value of the negative inertia is (a) $M = -50.752$ kg, (b) $M = -44.416$ kg, (c) $M = -40.256$ kg. The analytical values of the frequencies are (a) $f = 71.65$ Hz, (b) $f = 95.33$ Hz, (c) $f = 122.33$ Hz, which are very close to the numerical outcomes. The other parameters used in the computations are $h = 0.005$ m, $\rho = 2700$ kg/m$^3$, $D = 818.3$ Nm, $R = 1$ m, $N = 64$ and the side length of the square domain is $L = 10$ m.  The computations have been performed in \emph{Comsol Multiphysics} (version 5.2a).}
\label{FigureExpLocalisedModes}
\end{figure}
%%%%%%%%%%%%%%%%%%%%%%%%%%%%%%%%%%%%%%%%%%%%

Using the normalisation, we obtain
\begin{equation}\label{DisplacementPlateNormalisedResonators}
\tilde{\beta}^{4} \int_{0}^{\pi} \cos{(k \theta)} G\left(\sin{\left(\frac{\theta}{2}\right)}; \tilde{\beta} \right) \mathrm{d}\theta + 4 \pi \rho h \frac{R^{2} (\mathcal{C} - \mathcal{M} \omega^{2})}{\mathcal{C} \mathcal{M}} = 0 \, .
\end{equation}
After determining $\tilde{\beta}$ from \eqref{ZeroImaginaryPart}, $R$, $\mathcal{C}$ and $\mathcal{M}$ are chosen so that
\begin{equation}\label{ZeroRealPartResonators}
\frac{\tilde{\beta}^{2}}{4}\mathrm{I}_{n}\left(\frac{\tilde{\beta}}{2}\right) \mathrm{K}_{n}\left(\frac{\tilde{\beta}}{2}\right) + 4 \pi \rho h \frac{R^{2}(\mathcal{C} - \mathcal{M} \omega^{2})}{\mathcal{C} \mathcal{M}} = 0 \, .
\end{equation}
When $\tilde{\beta} \gg 1$, we have
\begin{equation}\label{BetaTildeLocalisedModeResonators}
\tilde{\beta} + 16 \pi \rho h \frac{R^{2}}{\mathcal{C} \mathcal{M}} \left( \mathcal{C} - \mathcal{M} \frac{D \tilde{\beta}^{4}}{16 R^{4} \rho h} \right) = 0 \, .
\end{equation}

\begin{figure}[h]%[!htcb]
\centering
\includegraphics[width=0.55\columnwidth]{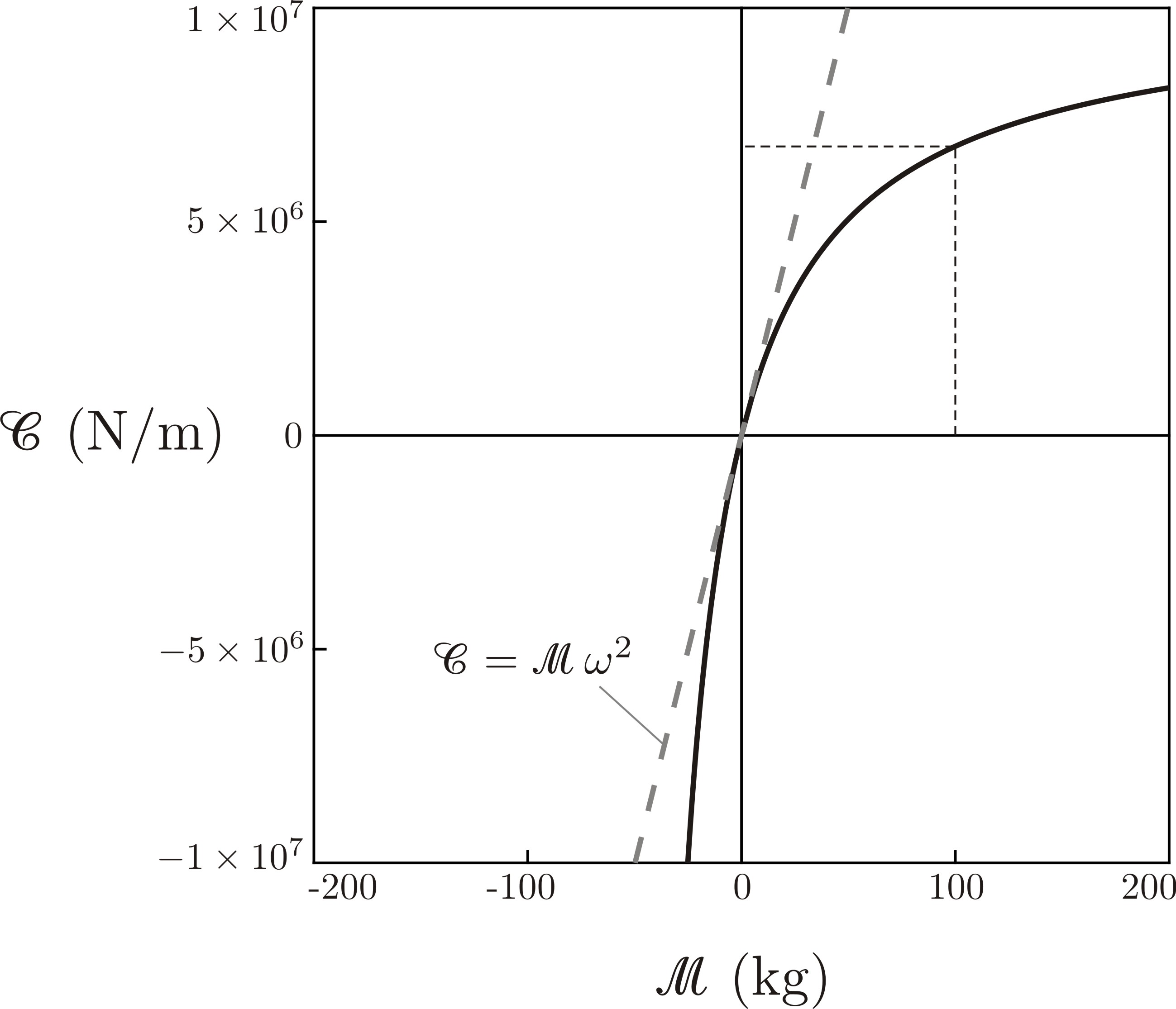}
\caption{\footnotesize Total stiffness $\mathcal{C}$ to be assigned to the resonators given their total mass $\mathcal{M}$ and the frequency parameter $\tilde{\beta}$. The dashed line is given by $\mathcal{C} = \mathcal{M} \, \omega^2$. The solid line is obtained for $M = -50.752$ kg, which is consistent with the case presented in Fig. \ref{FigureExpLocalisedModes}a. The figure shows that if the total mass of the resonators is taken as $\mathcal{M} = 100$ kg, the total stiffness must be equal to $\mathcal{C} = 6.764 \times 10^6$ N/m to give the negative inertia $M = -50.752$ kg.}
\label{FigureDeterminationValuesResonators}
\end{figure}

Spring-mass resonators can deliver the required negative inertia, which provides a larger range of admissible parameters to simulate exponentially localised vibration modes along the structured ring.
Fig. \ref{FigureDeterminationValuesResonators} shows how to choose the stiffness and the mass of the spring-mass resonator system to achieve the required negative inertia $M$. Namely,
\begin{equation} \label{calCM}
M = \frac{\mathcal{C} \mathcal{M}}{\mathcal{C} - \mathcal{M} \frac{D \tilde{\beta}^4}{16 R^4 \rho h}},
\end{equation}
where $\mathcal{C}$ and $\mathcal{M}$ represent the total stiffness and the total mass of the cluster of resonators; the other parameters are the same as in the previous sections. Given the required negative inertia $M$ and the value of the spectral parameter $\tilde{\beta}$, Fig. \ref{FigureDeterminationValuesResonators} shows $\mathcal{C}$ versus $\mathcal{M}$ according to the formula (\ref{calCM}).

%%%%%%%%%%%%%%%%%%%%%%%%%%%%%%%%%%%%%%%%%%%%
%%%%%%%%%%%%%%%%%%%%%%%%%%%%%%%%%%%%%%%%%%%%

\subsection{Exponentially localised waveforms}
\label{IllustrationsResonatorsSection}

Exponentially localised waveform solutions in a ring of spring-mass resonators are shown in Fig. \ref{FigureExpLocalisedModes}. In these computations, we have considered negative inertia, as discussed above. Strong localisation is observed in this case, as compared to Fig. \ref{FigureLeakyModes}.

The spring-mass resonators are tuned according to (\ref{calCM}) to choose the appropriate mass ${\mathcal{M}}$ and the stiffness ${\mathcal{C}}$ to achieve the required negative inertia $M$. For the computations of Fig. \ref{FigureExpLocalisedModes}, the range of admissible parameters ${\mathcal{M}}$ and ${\mathcal{C}}$ is shown in Fig. \ref{FigureDeterminationValuesResonators}, where the formula (\ref{calCM}) has been implemented numerically.

\section{Infinite grating of masses}
\label{InfiniteGratingSection}

\begin{figure}[h]%[!htcb]
\centering
\includegraphics[width=0.65\columnwidth]{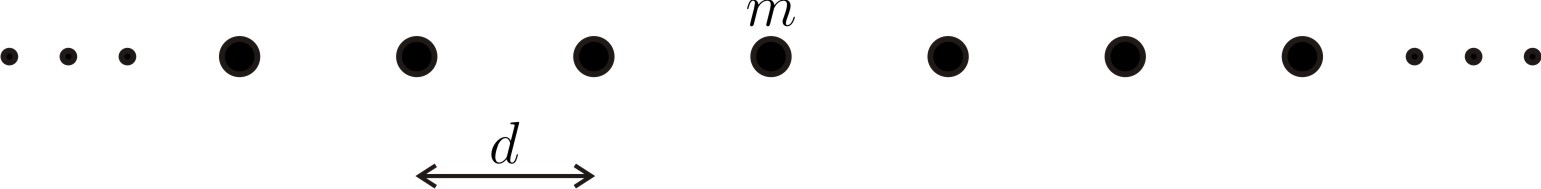}
\caption{\footnotesize Infinite grating of masses attached to an elastic plate.}
\label{FigureModelGrating}
\end{figure}

We show that the homogenised model can also be developed to identify trapped ``waveguide modes'' for an infinite grating of point masses $m$ embedded in an infinite elastic plate, as shown in Fig. \ref{FigureModelGrating}. These are often referred to as Rayleigh-Bloch waves \cite{eandp2007}.
Although it is tempting to think of this grating as the limit case of a circular ring as $R \to \infty$, we show that the trapped waves observed here cannot be obtained as the limit of the localised waveforms studied in Section \ref{HomogenisationSection}.

%%%%%%%%%%%%%%%%%%%%%%%%%%%%%%%%%%%%%%%%%%%%
%%%%%%%%%%%%%%%%%%%%%%%%%%%%%%%%%%%%%%%%%%%%

The system of equations is now given by
\begin{equation}\label{DisplacementPlateArray}
\sum_{k=-\infty}^\infty u\left(\bm{a}^{(k)}\right) G\left(\left|\bm{a}^{(k)}-\bm{a}^{(p)}\right|; \beta \right) d + \frac{d D}{m \omega^{2}} u\left(\bm{a}^{(p)}\right) = 0 \, ,
\end{equation}
where $d$ denotes the distance between the masses.
We note that $\left|\bm{a}^{(k)}-\bm{a}^{(p)}\right| = \left|k-p\right|d$. We seek the solution $u$ in the form $u\left(\bm{a}^{(k)}\right) = U \mathrm{e}^{\mathrm{i} \alpha k d}$, where $\alpha$ is the Bloch parameter. Then, Eq. (\ref{DisplacementPlateArray}) becomes
\begin{equation}\label{DisplacementPlateArray2}
\sum_{k=-\infty}^\infty \mathrm{e}^{\mathrm{i} \alpha d (k-p)} G\left(\left|k-p\right|d; \beta \right) d + \frac{D}{\mu \omega^{2}} = 0 \, ,
\end{equation}
where the mass per unit length $\mu = m/d$ has been introduced. The sum on the left-hand side of (\ref{DisplacementPlateArray2}) can be approximated by an integral, such that
\begin{equation}\label{DisplacementPlateArray3}
2 \int_{0}^{\infty} \cos{(\alpha s)} G(s;\beta) \mathrm{d}s + \frac{D}{\mu \omega^{2}} = 0 \, .
\end{equation}
Compared to Section \ref{HomogenisationSection}, this now is no longer the limit as $N \to \infty$, as $N$ has no meaning for the infinite array, but is instead the limit as $d \to 0$ and $m/d = O(1)$.

For real and positive values of $\alpha$ and $\beta$, Eq. (\ref{DisplacementPlateArray3}) takes the form
\begin{equation}\label{DisplacementPlateArray4}
\frac{1}{4 \beta^2} \left( \frac{1}{\sqrt{\alpha^2+\beta^2}} - \frac{\mathrm{i}}{\sqrt{\beta^2-\alpha^2}} \right) + \frac{D}{\mu \omega^{2}} = 0 \, .
\end{equation}
Assuming that $\alpha > \beta > 0$ and using (\ref{betom}), we obtain
\begin{equation}\label{DisplacementPlateArray5}
\frac{1}{4 \beta^2} \left( \frac{1}{\sqrt{\alpha^2+\beta^2}} - \frac{\mathrm{1}}{\sqrt{\alpha^2-\beta^2}} \right) + \frac{\rho h}{\mu \beta^{4}} = 0
\end{equation}
or, equivalently,
\begin{equation}\label{DisplacementPlateArray6}
\beta^2 \left( \frac{\mathrm{1}}{\sqrt{\alpha^2-\beta^2}} - \frac{1}{\sqrt{\alpha^2+\beta^2}} \right) = \frac{4 \rho h}{\mu} \, .
\end{equation}
This equation can be compared to (6.3) and (4.9) of \cite{eandp2007}. We observe that the homogenisation approximation extracts the $n=0$ term from (4.9), and then Eqs. (6.3) of \cite{eandp2007} and (\ref{DisplacementPlateArray6}) are consistent with each other.

At low frequencies, the dispersion relation (\ref{DisplacementPlateArray6}) leads to the following asymptotic approximation for the radian frequency $\omega$ as a function of the wavenumber $\alpha$:
\begin{equation}\label{DisplacementPlateArray7}
\omega \sim 2 \sqrt{\frac{D}{\mu}} \alpha^{3/2} \, ,
\end{equation}
when $\alpha > 4 \rho h/\mu$.

%%%%%%%%%%%%%%%%%%%%%%%%%%%%%%%%%%%%%%%%%%%%
\begin{figure}[h]%[!htcb]
\centering
\includegraphics[width=0.55\columnwidth]{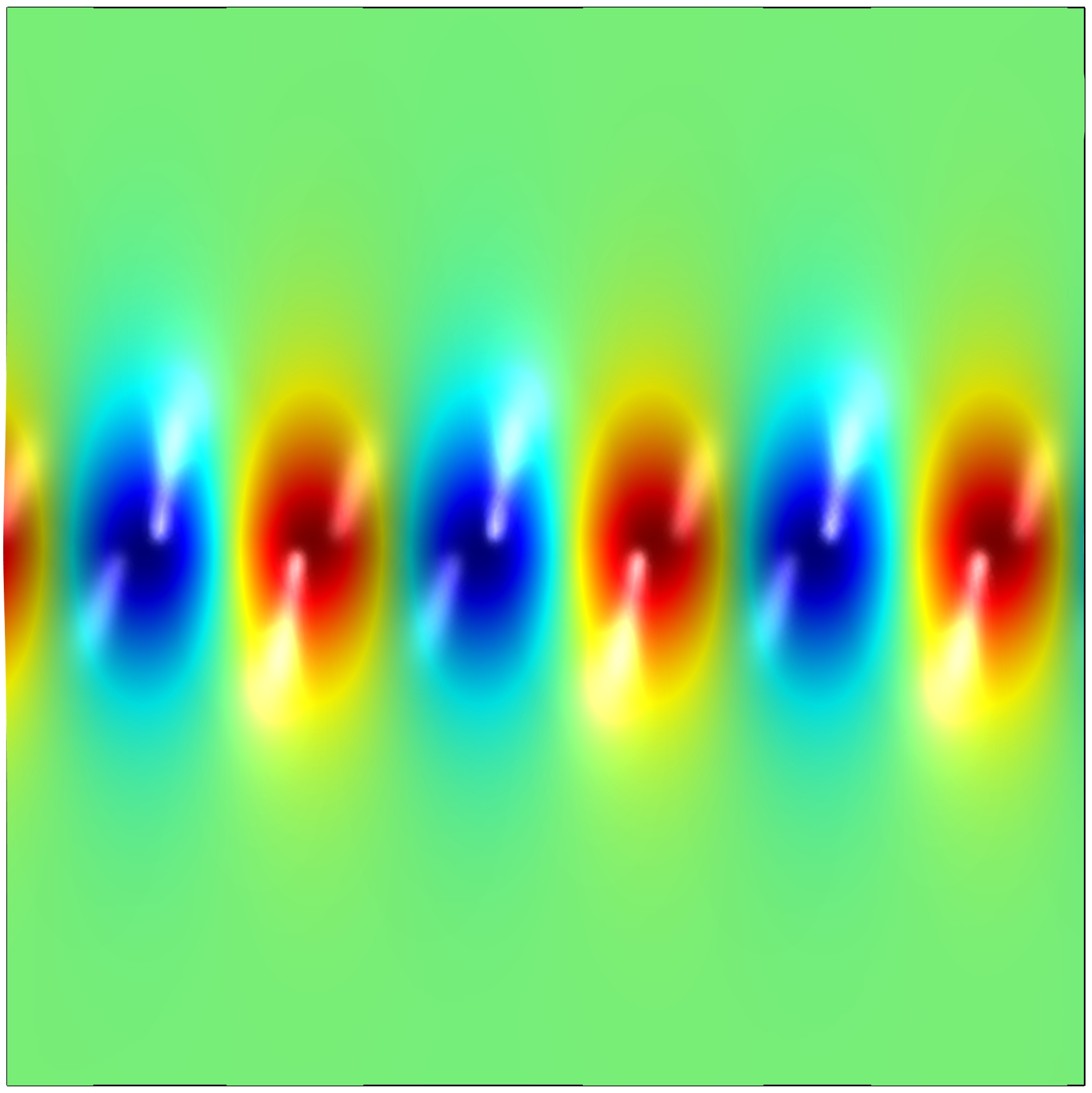}
\caption{\footnotesize Localised wave in an infinite array of masses connected to an elastic plate. In this computation, the wavenumber is $\alpha = 2$ 1/m and the radian frequency determined numerically is $\omega = 14.4$ rad/s, which is very close to the analytical value ($\omega = 15.7$ rad/s). The other parameters are $h = 0.005$ m, $\rho = 2700$ kg/m$^3$, $D = 818.3$ Nm, $m = 21.6$ kg, $d = 0.2$ m and $L = 10$ m.  The computation has been performed in \emph{Comsol Multiphysics} (version 5.2a).}
\label{FigureArray}
\end{figure}
%%%%%%%%%%%%%%%%%%%%%%%%%%%%%%%%%%%%%%%%%%%%

An example of a trapped flexural waveform for the infinite grating in the elastic plate is shown in Fig. \ref{FigureArray}, where quasi-periodic boundary conditions have been imposed on the vertical sides of the square domain. In this example, the wavenumber is chosen to be $\alpha = 2$ 1/m, which corresponds to the numerical value of the frequency $f = 5.3$ Hz. We note that this value of the frequency is close to the analytically predicted value, that is $f = 5.5$ Hz.
%which corresponds to the numerical value of the radian frequency $\omega = 14.4$ rad/s. We note that this value of the radian frequency is close to the analytically predicted value, that is $\omega = 15.7$ rad/s.

We point out that equations (\ref{DisplacementPlateArray3}) and (\ref{DisplacementPlateArray4}) do not have real and positive solutions $\alpha, \omega(\alpha)$ such that $\omega'(\alpha)=0$, i.e. the homogenisation approximation does not capture localised standing waves supported by the infinite grating of point masses in the elastic plate. This statement also holds true if point masses are replaced by spring-mass resonators. We also note that localised standing waves can be observed when the wavelength is comparable with $d$, which is outside the framework of the homogenisation approximation. This is described in Section 6 of \cite{eandp2007} and illustrated in Fig. 5a of the same paper.

\section{Discussion and concluding remarks}
\label{ConcludingSection}

We have presented a homogenisation model, which leads to closed form solutions that correspond to exponentially localised waveforms in structured plates. The cases considered here include flexural plates containing structured rings as well as infinite gratings.
%It has also been shown that localised modes in the infinite grating cannot be obtained in the limit of the exponentially localised waveforms identified in the structured ring.

Compared to the leaky waves, studied in detail in \cite{Hugo2016}, we have obtained exponentially localised ring-shaped waveforms by identifying the regime of negative inertia, which can be achieved by implementing specially tuned spring-mass resonators, as discussed in Section \ref{ResonatorsSection}. It has also been derived that a non-unique choice of resonator parameters (as shown in formula (\ref{calCM})) can be used to achieve the same value of the effective negative inertia.

The work has interesting extensions to the design of energy absorbers in flexural plates and construction of flexural elastic waveguides.

\section*{Acknowledgments}
\noindent
A.B.M., G.C. and R.V.C. would like to thank the EPSRC (UK) for its support through Programme Grant no. EP/L024926/1. R.C.M. gratefully acknowledges support for travel to Liverpool from the same grant.

%\section*{References}

\end{document}